\documentclass[twocolumn,showpacs,preprintnumbers,amssymb,amsmath]{revtex4}
\usepackage{graphicx}
\usepackage{dcolumn}
\usepackage{bm}

\newcommand{\bi}{\begin{itemize}}
\newcommand{\ei}{\end{itemize}}
\newcommand{\be}{\begin{equation}}
\newcommand{\ee}{\end{equation}}
\newcommand{\ba}{\begin{eqnarray}}
\newcommand{\ea}{\end{eqnarray}}
\newcommand{\bse}{\begin{subequations}}
\newcommand{\ese}{\end{subequations}}
\newcommand{\M}{{\cal {M}}}

\newcommand{\CP}{{\cal {P}}}
\newcommand{\la}{\langle}
\newcommand{\ra}{\rangle}
\newcommand{\kB}{k_{_B}}
\newcommand{\Vrot}{V_{\textrm{\tiny{rot}}}}
\newcommand{\rvir}{r_{\textrm{\tiny{vir}}}}
\newcommand{\Mvir}{M_{\textrm{\tiny{vir}}}}

\newcommand{\rhoNFW}{\rho_{\textrm{\tiny{NFW}}}}
\newcommand{\MNFW}{M_{\textrm{\tiny{NFW}}}}
\newcommand{\PhiNFW}{\Phi_{\textrm{\tiny{NFW}}}}
\newcommand{\dd}{\textrm{d}}

\begin{document}

\preprint{APS/123-QED}

\title{The spacetime associated with galactic dark matter halos.}

\author{Tonatiuh Matos$^\dagger$, Dar\'\i o N\'u\~nez$^\ddagger$ and Roberto A
Sussman$^\ddagger$}


\affiliation{
$^\dagger$ Departamento de F{\'\i}sica,
Centro de Investigaci\'on y de Estudios Avanzados del IPN, A.P. 14-740,
07000 M\'exico D.F., M\'exico.\\$^\ddagger$Instituto de Ciencias Nucleares
Universidad Nacional Aut\'onoma de M\'exico \\
A. P. 70-543,  M\'exico 04510 D.F., M\'exico\\}


\begin{abstract}
We show how an adequate post--Newtonian generalization can be obtained for
Newtonian dark matter halos associated with an empiric density profile.
Applying this approach to halos that follow from the well known numerical
simulations of Navarro, Frenk and White (NFW),  we derive all dynamical
variables and show that NFW halos approximatelly follow an ideal gas type of
equation of state which fits very well to a polytropic relation in the region
outside the core. This fact suggests that ``outer'' regions of NFW
halos might be related to equilibrium states in the
non--extensive Statistical Mechanics formalism proposed by Tsallis.
\end{abstract}


\maketitle

\section{Introduction}

The issue of dark matter clustering in halos of virialized galactic
structures is one of the most interesting open problems in astrophysics and
cosmology \cite{KoTu,Padma1,Peac,Ellis,Fornengo}. The physical properties of
this dark matter are uncertain, leading to various proposed physical matter
models: thermal sources, meaning a colissionless gas of weakly interacting
massive particles (WIMP's), which can be very massive ($m\sim 100-200$ GeV)
supersymmetric~\cite{Ellis} (``cold dark matter'' CDM) or self--interacting
less massive ($m\sim$ KeV) particles~\cite{scdm,wdm} (``warm'' DM). Other
proposals include scalar fields (real and complex)~\cite{sfe_1,sfe_2}, global
momopoles~\cite{sfe_3}, axions, etc. However, all these models must comply
from inferred direct and indirect observations that reveal the presence of DM:
velocity profiles of rotating stars, microlensing and tidal effects affecting
satelite galaxies and galaxies within galactic clusters. Galactic DM is mixed
with visible baryonic matter (stars and gas) clustering in galactic disks,
making up about 5--10 \% of the total galactic mass. Hence, it is a good
approximation to identify the gravitational field of a galaxy with that of
its DM halo, considering visible matter as ``test particles'' in this
field~\cite{HGDM, Lake}.

Realistic galactic halos are obviously not spherically symmetric,
but they are approximatelly so, since their global rotation is not
dynamically significant~\cite{urc}. Hence, we will consider throughout this
paper that halos are spherically symmetric equilibrium configurations.
Assuming the CDM paradigm and spherical symmetry, we can distinguish two
types of halo models: those obtained from a Kinetic Theory approach, whether
based on specific theoretical considerations or on convenient ansatzes that
fix a distribution function satisfying Vlassov's equation~\cite{BT}, or those
emerging from ``universal'' density profiles obtained empirically by N--body
numerical simmulations~\cite{nbody_1,nbody_2,nbody_3}. In this paper we will
consider the latter apprach, based on the well known numerical simulations of
Navarro, Frenk and White (NFW)~\cite{nbody_1,LoMa}. Although these simulations
yield virialized equilibrium structures that reasonably fit CDM structures at
a cosmological scale ($\agt 100$ Mpc), some of their predictions in smaller
scales (``cuspy'' density profiles and excess substructure) seem to be at
odds with observations~\cite{cdm_problems_1,cdm_problems_2}, especially
those based on galaxies with low surface brightness (LSB), which are supposed
to be overwhelmingly dominated by DM and so well suited to examine the
predictions of various DM models.

 Galactic halos are newtonian systems
characterized by typical velocities, ranging from $5-10$ km/sec for dwarf
galaxies up to about $1000-3000$ km/sec for rich clusters. However, we believe
that a study of these systems under General Relativity, as a
post--Newtonian approximation, might provide new information that can be
useful and interesting for gravitational lensing and for the study of the
interplay between cosmological scale evolution and galactic DM. In any case,
since General Relativity is the best available theory of classical
gravity, it is relevant from a theoretical point of view to be able to
construct  spacetimes that are suitable for important self--gravitating
structures like galactic halos.

All DM halo models derive a full set of dynamical variables from a given
``mass--density'' profile. In a post--newtonian relativistic generalization
we will assume that this density is the dominant rest--mass contribution to
the matter--energy density, made up by rest--mass and an internal energy
term proportional to suitably defined temperature and pressure (of a kinetic
nature). Thus, we will assume an ``ideal gas'' type of equation of
state~\cite{RKT,Padma2,Padma3} in which this internal energy density becomes
determined only by the hydrostatic equations themselves in the case of
isotropic velocity distributions. In the anistropic case, which we leave for
a future paper~\cite{enproceso}, various empirical ansatzes can be assumed in
order to relate radial and angular components of the stress tensor.

\section{Newtonian NFW galactic halos.}

Following the CDM paradigm and assuming spherical symmetry, galactic halos
must satisfy the following Newtonian equations of hydrostatic equilibrium
\ba M\,' = 4\,\pi\,\rho\,r^2,\label{Mr}\\
\Phi\,' = \frac{ 4\,\pi\,G M}{r^2},\label{Phir}\ea
as well as the Navier--Stokes equation
\begin{equation}P\,' = -\rho\,\Phi\,'-\frac{2\,\alpha}{r}\,P,
\label{Pr}\end{equation}
where $P$ is the ``radial'' pressure and
\begin{equation}\alpha = \frac{P-P_\perp}{P},\label{alpha} \end{equation}
is the anisotropy factor relating $P$ with the tangential pressure
$P_\perp$. Since we have three equations for five unknowns
($\rho,\,M,\,\Phi,\,P,\,\alpha$), this system can be made determined if
two of these five functions becomes specified, for example, by assuming an
``equation of state'' somehow relating $P$ and
$P_\perp$ with $\rho$. In the case of N--body numerical simmulations, we
have virialized structures whose density profile $\rho=\rho(r)$ can be
approximately fit to a ``universal'' empirical
function~\cite{urc,nbody_1,LoMa}. The Newtonian system
(\ref{Mr})--(\ref{alpha}) becomes determined once we have this density
profile together with a suitable expression for $\alpha$. In general, the
simmulations yield anisotropic velocity  distributions, so that specific
ansatzes can be assumed or prescribed~\cite{OM} for $\alpha\ne 0$.

The well known N--body numerical simmulations by Navarro, Frenk and White
(NFW) yield the following ``universal'' expression for the density profile of
virialized galactic halo structures~\cite{nbody_1,LoMa}:
\begin{equation}
\rho_{_{\mathrm{NFW}}}=\frac{\delta_0\,\rho_0}{x\,\left(1+x \right)^2},
\label{rho_NFW}
\end{equation}
where
\begin{eqnarray}
\delta_0 &=& \frac{\Delta\,c_0^3}{3\left[\ln\,(1+c_0)-c_0/ (1+c_0)\right]},
\label{delta0} \\
\nonumber \\
\rho_0&=&\rho_{\mathrm{crit}}\,\Omega_0\,h^2 = 253.8 \, h^2\,\Omega_0\,
\frac{M_\odot}{\text{kpc}^3},  \label{rho0} \\
x &=& \frac{r}{r_s},\qquad r_s = \frac{\rvir}{c_0},  \label{x}
\end{eqnarray}
The virial radius $\rvir$ is given in terms of the virial mass $\Mvir$ by the
condition that average halo density equals $\Delta$ times the cosmological
density $\rho_0$~\cite{Padma1}
\begin{equation}
\Delta\, \rho_0 = \frac{4\,\pi\, \Mvir}{3\,\rvir^3},  \label{rvir}
\end{equation}
where $\Delta$ is a model--dependent numerical factor (for a $\Lambda$CDM
model with total $\Omega_0=1$ we have $\Delta\sim 100$~\cite{LoHo}). The
concentration parameter $c_0$ can be expressed in terms of $\Mvir$
by~\cite{c0}
\begin{equation}
c_0 = 62.1 \times
\left(\frac{\Mvir\,h}{M_\odot}\right)^{-0.06}\,\left(1+\epsilon\right),
\label{c0}
\end{equation}
where $-0.5\alt \epsilon\alt 0.5 $.
Hence all quantities depend on a single free parameter $\Mvir$ with a
dispersion range given by $\epsilon$ for different halo concentrations. The
NFW mass function and Newtonian potential follow from integrating (\ref{Mr})
and (\ref{Phir}) for $\rho$ given by (\ref{rho_NFW})
\begin{eqnarray}
\MNFW &=& 4\,\pi\,r_s^3\,\delta_0\,\rho_0\,\left[\ln(1+x)
-\frac{x}{1+x}\right],  \label{M_NFW}\\
\PhiNFW &=& -V_0^2\,\frac{\ln(1+x)}{x}
\end{eqnarray}
complying with the boundary conditions
\ba\MNFW(0) &=& 0,\qquad \MNFW(\rvir)=\Mvir,\label{nbcs}\\
-\PhiNFW(0) &=& V_0^2 \ = \ 4\,\pi\, G\,
\delta_0\,\rho_0\,r_s^2,\label{V0}  \ea
while circular rotation velocity (normalized by the characteristic velocity
$V_0^2$) is simply
\begin{equation}\Vrot^2 = r\,\Phi'=\frac{4\pi G M}{r} =
V_0^2\left[\frac{\ln(1+x)}{x}-\frac{1}{1+x}\right].\label{Vrot}\end{equation}
Given $\rhoNFW,\,\MNFW$ and $\PhiNFW$, radial and tangential pressures
follow from integrating (\ref{Pr}) for a specific choice of
$\alpha$. There are analytic solutions of (\ref{Pr}) for $\alpha=0$
(isotropic case) and for various empiric expressions for $\alpha$. A thorugh
Newtonian treatement of NFW halos is found in~\cite{LoMa}

Notice that according to the density profile (\ref{rho_NFW}) we have a
diverging density at the symmetry center ($x=0$). This is obviously an
unphysical feature and points out to the fact that (\ref{rho_NFW})
has not been derived from any theoretical argumentation, but is simply a
convenient empirical formula that fits the outcome of the NFW numerical
simulations which show that $\rhoNFW\propto 1/x$ near the central halo
region. Since the virial radius $\rvir$ is the physical size of the
resulting halos and these simulations are unable to provide adequate
resolution for distances to the halo center smaller than approximately 1\% of
the virial radius~\cite{NSres}, all halo quantities presented here are,
strictly speaking, only valid between a minimal $r\sim 0.01 \,\rvir$ and
$\rvir$ (see section VII for further discussion on this issue).

\section{The spacetimes of galactic halos}

As a good approximation \cite{otro}, we can consider the metric of
a galactic halo to be given by a suitable ``weak field'' limit of
the spherically symmetric static line element
\ba \dd s^2 \ = \ -\exp\left(\frac{\Phi}{c^2}\right)\,c^2\,\dd
t^2+\frac{\dd
r^2}{1-\kappa_0\,M/r},\nonumber\\
+r^2(\dd\theta^2+\sin^2\theta\dd\phi^2),\label{metric}\ea
where $\Phi(r)$ is the relativistic generalization of the Newtonian
gravitational potential and $\kappa_0=2 G/c^2$, so that
$M(r)$ has mass units. We will assume a momentum--energy tensor of the
form
\ba T^{ab}  = \ \mu\,u^a\,u^b + p\,h^{ab} + \Pi^{ab},\label{Tab}\ea
where $u^a = \exp(-\Phi/c^2)\,\delta^a\,_t$, \ $h^{ab}  =  g^{ab}+u^a\,u^b$
and
$\Pi^{ab}$ is the anisotropic and traceless ($\Pi^a\,_a=0$) stress tensor,
which for the metric (\ref{metric}) takes the form
\ba \Pi^a\,_b \ = \ \textbf{diag}\,[0,\,-2\Pi,\,\Pi,\,\Pi]\ea
so that $p$ and $\Pi$ relate to $P$ and $P_\perp$ by
\begin{equation}P_\perp-P \ = \ 3\,\Pi,\qquad 2\,P_\perp+P \ = \
3\,p.\label{pps}\end{equation}
The field equations and momentum balance ($T^{ab}\,_{;b}=0$) associated with
(\ref{metric})-(\ref{pps}) are
\ba M\,' &&= \ 4\,\pi\,\mu\,r^2/c^2,\label{M2r}\\
\Phi\,' &&= \
\frac{\kappa_0\,c^2}{2}\,\frac{M+4\,\pi\,P\,r^3/c^2}{r\,(r-\kappa_0\,M)},
\label{Phi2r}\\ P\,'&& = \
-(\mu+P)\,\frac{\Phi\,'}{c^2}-\frac{2\,\alpha}{r}\,P,\label{P2r}\ea
where $\alpha$ is given by (\ref{alpha}). These equations are
the relativistic generalization of (\ref{Mr})--(\ref{Pr}), though we
must provide a relation between $\mu$ and $\rho$. Since the
particles in the collisionless gas making up galactic halos are interacting
very weakly and the velocity anisotropies tend to be small: $0\leq
\alpha\alt 0.2$~\cite{LoMa}, it is reasonable to assume that it is nearly an
ideal gas and that total matter--energy density,
$\mu$, is the sum of a dominant contribution from rest--mass density,
$\rho\,c^2$, and an internal energy term that is proportional to the pressure
$P$ and to the velocity dispersion $\sigma^2=\la v^2\ra\simeq \la
v_\perp^2\ra$. Hence it is reasonable to assume the equation of state of a
non--relativistic (but non--Newtonian) ideal gas~\cite{RKT}
\begin{equation} \mu \ = \
\rho\,c^2\,\left[1+\frac{3}{2}\,\frac{\sigma^2}{c^2}\right],\qquad P
\ = \ \rho\,\sigma^2,\label{NRIGES}\end{equation}
where the velocity dispersion is related to a kinetic temperature $T$ by
\begin{equation}\sigma^2 \ = \ \frac{P}{\rho} \ = \
\frac{\kB\,T}{m},\label{sigma2}\end{equation}
Since characteristic velocities in galactic halos are
Newtonian, we have $\sigma^2/c^2\ll 1$ and $\mu\approx
\rho\,c^2$ and so $P\simeq P_\perp\ll \rho\,c^2 $, so that
(\ref{NRIGES}) provides a plausible equation of state for a
relativistic generalization of galactic halos. It is evident also that in
the newtonian limit $\sigma^2/c^2\ll 1$ we recover the equilibrium
equations (\ref{Mr})--(\ref{Pr}).

What needs to be done now is to insert the equation of state (\ref{NRIGES})
into the field equations (\ref{M2r})--(\ref{P2r}), which becomes a set of
equations that, just like (\ref{Mr})--(\ref{Pr}), becomes determined once we
specify the functional relation $\rho=\rho(r)$ or $\rho=\rho(\Phi)$ and
$\alpha(r)$. However, we do not need all the three equations
(\ref{M2r})--(\ref{P2r}), since numerical simulations yield the density
profile (\ref{rho_NFW}), we will assume
\begin{equation} \rho(r) \ = \ \rhoNFW(r)\end{equation}
and eliminate $\Phi'$ from (\ref{Phi2r}) and (\ref{P2r}), leading to
the following set:
\ba M\,' &&= \
4\,\pi\,\left[\,\rhoNFW+\frac{3}{2}\,\frac{P}{c^2}\right]r^2,\label{M3r}\\
 P\,'&& = \
-\frac{\kappa_0c^2}{2}\,\frac{[\rhoNFW+\frac{5}{2}P/c^2]\,
\,[M+4\,\pi\,P\,r^3/c^2]} {r\,(r-\kappa_0\,M)}\nonumber\\ && \quad \
-\frac{2\,\alpha}{r}\,P,\label{P3r}\ea
which becomes fully determined once we know $\alpha(r)$.
We will solve these equations in a post--Newtonian
approximation by keeping only terms up to order
$\sigma^2/c^2$. The velocity dispersion $\sigma$ (and/or $T$) can be obtained
afterwards from $P$ through (\ref{NRIGES}) and (\ref{sigma2}).

As mentioned before, we have $\rhoNFW\to\infty$ as
$x\to 0$. A quick calculation of the Ricci scalar using (\ref{Phi2r}),
(\ref{M3r}) and (\ref{P3r})
\begin{equation}R \ = \ \frac{8\pi
G}{c^4}\,\left[\rhoNFW\,c^2+\left(2\,\alpha-\frac{3}{2}\right)\,P\right],
\label{Ricci1}
\end{equation}
reveals the existence of a curvature singularity as $r\to 0$. This is wholy
unphysical, since NFW halos within a general relativistic treatment should be
weak field static spacetimes. However, as mentioned in the
last parragraph of section II, all variables associated with NFW simulations
are valid in the range $0.01 \alt r/\rvir
\leq 1$. In section VII we discuss how this issue can be dealt with
appropriately.

\section{Post--Newtonian equations for NFW halos.}

It is convenient to work with the following adimensional variables
\ba Y && = \ \frac{\rhoNFW}{\delta_0\,\rho_0} \ = \
\frac{1}{x\,[1+x]^2},\label{Y}
\\
\M && = \ \frac{M}{4\,\pi\,\delta_0\,\rho_0\,r_s^3} \ = \
\frac{c_0^3\,\Delta\,M}{3\,\delta_0\,\Mvir},\label{CM}\\
\CP && = \ \frac{P}{\delta_0\,\rho_0\,V_0^2},\label{eq:CP}\ea
where the structural parameters $\delta_0,\,\rho_0,\,c_0,\,\Delta$ and
$V_0^2$ have been introduced in section II. The field equations
(\ref{M3r})--(\ref{P3r}) now becomes
\ba \frac{\dd \M}{\dd x}  &&= \
\left[Y+\frac{3}{2}\,\varepsilon\,\CP\right]\,x^2,\label{Mx}\\
\frac{\dd \CP}{\dd x} &&= \
-\frac{\left[Y+\frac{5}{2}\,\varepsilon\,\CP\right]\,\left[\M+\varepsilon\,
\CP\,x^3\right]}{x\,\left[x-2\,\varepsilon\,\M\right]}+
\frac{2\alpha}{x}\,\CP,\nonumber\\
\label{Px}\ea
where
\begin{equation} \varepsilon \ = \ \frac{V_0^2}{c^2},
\label{epsilon}\end{equation}
so that in the limit $\varepsilon\to 0 $ we recover the Newtonian
equations (\ref{Mr}), (\ref{Phir}) and (\ref{Pr}). The system
(\ref{Mx})--(\ref{Px}) can be integrated by demanding that $\M$ and $\CP$ comply
with appropriate boundary and initial conditions. Since we have to use the
explicit form of $Y$ in (\ref{Y}), then the analytic or numerical solutions
of (\ref{Mx})--(\ref{Px}) for specific choices of
$\alpha$, boundary conditions depend on $\Mvir$ through the definitions
(\ref{delta0}) and (\ref{c0}).

The metric function $M=V_0^2\,r_s\,\M$ follows from (\ref{Mx}), while
$\Phi$ can be obtained by integrating
\begin{equation}\frac{\dd}{\dd y}\,\left(\frac{\Phi}{V_0^2}\right) \ =
\ \frac{\M+\varepsilon\,\CP\,y^3}{y\,[y-2\,\varepsilon\,\M]}. \label{Phix}
\end{equation}
The relativistic generalization of the
Newtonian rotation velocity profile are the velocities of test observers
along circular geodesics. From \cite{sfe_1,sfe_2,HGDM}, these velocities are
$V_{\textrm{\tiny{rot}}}^2=r\,\Phi'$, which in terms of the adimensional
variables becomes
\begin{equation}\frac{V_{\textrm{\tiny{rot}}}^2}{V_0^2} \ = \
\frac{\M+\varepsilon\,\CP\,y^3}{y-2\,\varepsilon\,
\M},\label{Vx}\end{equation}

Since $V_0$ for typical galactic halos ranges from a few km/sec to $\sim
3000$ km/sec, the post--Newtonian corrections of order $V_0^2/c^2$ will be
very small: between $O(\varepsilon)\sim 10^{-9}$ and $O(\varepsilon)\sim
10^{-6}$. The post--Newtonian system associated with
(\ref{Mx})--(\ref{Px}) can be given as
\ba \frac{\dd \M}{\dd x}  &=& \
Y\,x^2+O(\varepsilon),\label{Mx0}\\
\frac{\dd \CP}{\dd x} &=& \
-\frac{Y\,\M}{x^2}+
\frac{2\alpha}{x}\,\CP+O(\varepsilon),
\label{Px0}\ea
while (\ref{Phix}) and (\ref{Vx}) become
\ba\frac{\dd}{\dd y}\,\left(\frac{\Phi}{V_0^2}\right) \ = \
\frac{\M}{x^2}+O(\varepsilon),\label{PhiV}\\
\frac{V_{\textrm{\tiny{rot}}}^2}{V_0^2} \ = \
\frac{\M}{x}+O(\varepsilon),\label{VrotPN}
\ea

\section{Analytic solutions}

In the post--Newtonian equations given above, $\CP$ is decupled from $\M$
and $\Phi$, thus (irrespective of the form of $\CP$) the metric elements for
all NFW halo spacetimes are up to order $\varepsilon$
\ba
-g_{tt} &=& \textrm{e}^{2\Phi/c^2} \approx
1-\frac{2\,\ln(1+x)}{x}\,\varepsilon+O(\varepsilon^2),\\
 g_{rr} &=&
\left[1-\frac{2\,\varepsilon\,\M}{x}\right]^{-1}\nonumber\\ &\approx&
1+2\,\left[\frac{\ln(1+x)}{x}-\frac{1}{1+x}\right]\varepsilon
+O(\varepsilon^2),
\ea
where the metric functions $M$ and $\Phi$ obtained from
(\ref{Mx0}) and (\ref{PhiV}) comply with the boundary conditions
(\ref{nbcs}) and (\ref{V0}) (see also \cite{otro}). Notice that
$M$ and $\Phi$ are finite at the center, even if $Y$ diverges.
Also, even if $M$ diverges as $r\to\infty$, the metric components
shown above are well behaved asymptotically, tending to flat
spacetime:\, $-g_{tt}\to 1$ and $g_{rr}\to 1$ \, as $x\to\infty$.
However, the NFW spacetimes do not comply with the regularity at
the center of a spherically symetric spacetime which requires the
vanishing of all spacelike gradients (such as $M'$ and $\Phi'$).
In fact, the Ricci scalar (\ref{Ricci1}) in the post--Newtonain
limit becomes: \, $R=-2\,Y\,\varepsilon+O(\varepsilon^2)$, hence
there is a curvature singularity in the center even if the metric
functions do not diverge.

Even if all NFW halos have the same rest--mass density
$Y$ and metric functions $M,\,\Phi$, the form for the pressure depends on the
assumptions one might make about $\alpha$ and suitable boundary conditions.
For the remaining of this paper we will consider only the case of isotropic
velocity distributions, leading to $\alpha=0$. In this case, the
post--Newtonian Navier--Stokes equation (\ref{Px0}) has the following
analytic solution:
\begin{widetext}
\ba\CP &=&
C+\frac{3}{2}\,\left[\ln(1+x)\right]^2+A(x)\,\ln(1+x)-\frac{1}{2}\,
\ln\,x+3\,\textrm{dilog}\,(1+x)-B(x)+O(\varepsilon),\nonumber\\
 &&\qquad\qquad A(x)=\frac{1-3x+5x^2+x^3}{2x^2\,(1+x)},\qquad\qquad
B(x)=\frac{1+9x+7x^2}{2x\,(1+x)^2},\label{eq:CP1}\ea
\end{widetext}
where $C$ is a constant and the dilogarithmic function is defined as
\begin{equation}\textrm{dilog}(y) \ = \
\int_1^y{\frac{\ln t\,\dd t}{1-t}}.\nonumber\end{equation}
In order to determine $C$, we need to examine the boundary conditions
of $\CP$.

\section{Polytropic equation of state}

\begin{figure}
\centering
\includegraphics[height=9cm]{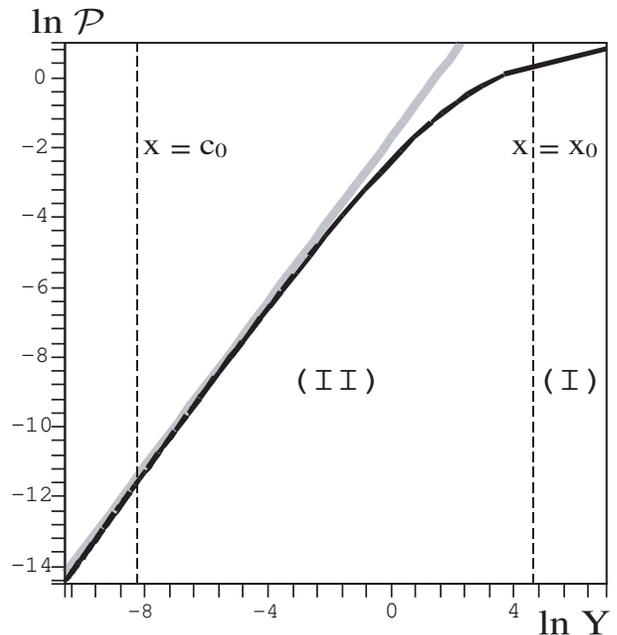}
%
%
\caption{The thick curve is the plot of $\ln\,\CP$ vs $\ln\,Y$, equivalent to
plotting $\ln\,P$ vs $\ln\,\rho\,V_0^2$. The radius $x=x_0\approx 0.01\,c_0$,
corresponding to the minimal resolution of numerical simulations, marks the
``inner'' region (I). The ``outer'' region (II) denotes the halo up to its
physical radius, the virial radius $x=c_0=15$. As a comparison we show a line
with slope $1.18$ (thick grey line) that would correspond to the poytropic
relation with $n\approx 5.5$. Notice how the NFW halo approximately fits this
relation, except near the center where the density profile becomes cuspy.}
\label{fig:1}       
\end{figure}

The asymptotic behavior ($x\gg 1$) of $\CP$ in (\ref{eq:CP1})
\begin{equation}\CP \approx
C-\frac{\pi^2}{2}+\frac{4\,\ln\,x-3}{16\,x^4}+O(x^{-5})\end{equation}
implies that an asymptotically flat configuration arises if we choose
$C=\pi^2/2$,\, so that $\CP\to 0$, scaling asymptotically as $\CP\propto
\ln\,x/x^4$. Since $Y$ scales asymptotically as $1/x^3$, this indicates a
sort of power law relation between $\CP$ and $Y$ that (at least
asymptotically) might be similar to a polytropic relation of the form
\begin{equation}\CP \approx K\,Y^{1+1/n},\label{poly}\end{equation}
where $K$ and $n$ (polytropic index) are constants. In order to examine the
functional relation between $Y$ and $\CP$, we provide in figure \ref{fig:1} a
logarithmic plot of $\CP$ vs $Y$ (or equivalently $\ln P$ vs
$\ln \rho\,V_0^2$), for the asymptotically flat case with
$C=\pi^2/2$ applied to a halo with $\Mvir =10^{12}\,M_\odot$, corresponding
to a virial radius marked by $x=c_0\sim 15$ ($\sim 150$ kpc). For theoretical
reference we show the curve associated with a polytropic
relation (\ref{poly}) with $n\approx 5.5$ and $K\approx \exp(-1.7)$. As shown
by the figure, the asymptotically flat NFW configuration fits very well this
polytrope, except for high density values corresponding to smaller $x$, up to
the value $x=x_0\approx 0.01\,c_0$ that marks the resolution limit of
numerical simulations ($\sim 1$ kpc). This behavior is reasonable, since
closer to the center ($x$ close to $x_0$) the NFW density profile becomes
cuspy, while polytropic density profiles are characterized by a ``flat
core''~\cite{BT}.

\section{Discussion and conclussion}

The fact that NFW halos asymptoticaly comply with a polytropic relation
with $n\approx 5.5$ is quite significant, since stellar polytropes
characterized by (\ref{poly}) are the equilibrium state associated with the
entropy functional in the non--extensive entropy formalism derived by
Tsallis and coworkers~\cite{PL,Tsallis,TS1, TS2}. In its application to
self--gravitating collisionless systems this formalism is characterized by
the free parameter $q=(2n-1)/(2n-3)$, so that the isothermal sphere
(equilibrium state for the usual Boltzmann--Gibbs entropy functional)
follows in the ``extensive entropy'' limit $n\to\infty$ (or $q\to 1$).
Assuming Tsallis theory to be correct, the empiric verification (see
Figure \ref{fig:1}) that NFW halos outside the ``inner'' core satisfy a
polytropic relation might indicate that in this ``outer'' region the NFW
numerical simulations yield self--gravitating configurations that approach an
equilibrium state characterized by the Tsallis parameter
$q\approx 1.25$. However, while the central cusps in the density profile that
are predicted by NFW simulations seem to be at odds with
observations~\cite{cdm_problems_1,cdm_problems_2}, there is no conflict
between these observations and the
$1/x^3$ scaling of the NFW density profile outside the core region (as well
as the rotation velocity profile from (\ref{VrotPN})). Although the issue of
the cuspy cores is still controversial, if galactic halos seem to exhibit
flat density cores, their profiles could be adjusted to stellar polytropes
and this might be helpful in providing a better empirical verification of
Tsallis' formalism. However, this idea must be handled with due case, since
stellar polytropes follow from an isotropic velocity distribution, while
galactic halos with such distributions could be unrealistic.

As pointed out before, the density profile of NFW halos diverges at the
center. Apparently this issue has not bothered astrophysicists, since (as
mentioned before) the cuspy cores of NFW numerical simulations are meant
to show a density scaling of $1/x$ near the center and these
simulations cannot resolve distances to the halo center smaller than 1\,\% of
the virial radius~\cite{NSres}. One way to deal with this unphysical feature,
leading to a better description of these halos, would be to perform a smooth
matching between NFW spacetimes and a small central region with a regular
density profile. An adequate radius for this ``inner'' region could be the
minimal resolution scale in numerical simulations ($x=x_0\sim 0.01\, c_0$).
Another improvement could be a smooth matching of the NFW spacetime to a
Schwarzschild vacuum exterior at the virial radius
$x=c_0$, which is the physical radius of the halo. One of the matching
conditions in this latter case would be $\CP(c_0)=0$, implying a different
choice of the integration constant $C$ in (\ref{eq:CP1}). Another necessary
improvement is the study of the anisotropic cases for which $\alpha\ne 0$.

We have constructed the spacetime corresponding to post--Newtonian
generalizations suitable to NFW halos. Although we have presented only the
idealized case with isotropic pressure, the methodology that we followed
here can be applied, in principle, to any Newtonian model of galactic
halos. We believe that it is necessary to study galactic halo models (NFW, as
well as other empiric or theoretical models) within a wider framework
including the usual thermodynamics of self--gravitation
systems~\cite{Padma2,Padma3}, as well as alternative approaches such as
Tsallis' formalism~\cite{Tsallis,PL,TS1,TS2}. Such an improvement and
extension of the present study of NFW halos are being pursued
elsewhere~\cite{enproceso}.

\section{acknowledgements}

It is a pleasure to participate with this work in the number
dedicated to honor our friend and college Prof. Alberto Garc\'\i
a. RAS acknowledges financial support from grant PAPIIT-DGAPA
number IN117803, DN does so from grant DGAPA-UNAM IN122002. TM
acknowledges partial financial support by CONACyT M\'exico, under
grants 32138-E and 34407-E.

\end{document}